# Synthesis, characterization and magnetic properties of room-temperature nanofluid ferromagnetic graphite

N. S. Souza[1], S. Sergeenkov[1,a)], C. Speglich[1], V. A. G. Rivera[1], C. A. Cardoso[1], H. Pardo[2], A. W. Mombrú[2], A.D. Rodrigues[1], O.F. de Lima[3] and F. M. Araújo-Moreira[1,b)]

[1]*Materials and Devices Group; Department of Physics and Physical Engineering, Universidade Federal de São Carlos, 13565-905 São Carlos, SP, Brazil*

[2]*Crystallography, Solid State and Materials Laboratory (Cryssmat-Lab), DEQUIFIM, Facultad de Química, Universidad de la República, P.O. Box 1157, CP 11800, Montevideo, Uruguay*

[3]*Instituto de Fisica "Gleb Wataghin", UNICAMP, 13083-970 Campinas, SP, Brazil*

**Abstract**

We report the chemical synthesis route, structural characterization, and physical properties of nanofluid magnetic graphite (NFMG) obtained from the previously synthesized bulk organic magnetic graphite (MG) by stabilizing the aqueous ferrofluid suspension with an addition of active cationic surfactant. The measured magnetization-field hysteresis curves along with the temperature dependence of magnetization confirmed room-temperature ferromagnetism in both MG and NFMG samples.

*Keywords*: Graphite; nanoparticles; suspensions; ferrofluids; magnetic properties.

PACS: 75.50Dd ; 75.50.Mm; 75.50.Tt; 81.05.Uw; 82.70.Kj

---

a) Corresponding author; e-mail address: sergei@df.ufscar.br

b) Research Leader; e-mail address: faraujo@df.ufscar.br

Nanofluids can be defined as fluids containing suspended solid particles with sizes in the nanometers scale. In recent years, substantial progress has been made in developing technologies in the field of magnetic microspheres, nanospheres and nanofluids.[1-4] A most recognizable class of magnetically controllable nanofluids (simultaneously exhibiting both *fluid* and *magnetic* properties) are suspended colloids of nano sized iron oxide particles ($Fe_3O_4$ or $\gamma$-$Fe_2O_3$). At the same time, the important for applications biocompatible ferrofluids normally use water as a vehicle. In order to prevent agglomeration, the magnetic nanoparticles have to be stabilized by ionic interaction using some kind of bioagent (like, e.g., fatty, aspartic and glutamic acids or peptides). Alternatively, the co-precipitation of ferrous/ferric ions can be performed in the presence of the appropriate biopolymer (such as polyvinyl alcohol or polyethylen glycol). Several clinically important enzymes and proteins (including, among others, bovine serum albumin, streptokinase, chymotrypsin, and glucose oxidase) have been immobilized based on this method. On the other hand, carbon materials constitute one of the most fascinating classes of structures, exhibiting a wide variety of forms and properties. They have been increasingly attracting the attention of the scientific community mainly because of their potential applications in *high-tech* devices. For these reasons, the availability of macroscopic quantities of bulk room-temperature magnetic carbon-like graphite is of utter importance not only for a wide number of natural sciences, but also for technological applications of this material in engineering (as well as in medicine and biology).[5-10] We have already reported the method to obtain macroscopic quantities of pure *bulk* ferromagnetic graphite.[11-14] The as-obtained modified graphite has a strong magnetic response even at room temperature (which manifests itself through a visible attraction by a commercial magnet).[12]

In this Letter, we present the chemical route for synthesizing nanofluid magnetic graphite (NFMG) as well as its structural and magnetic properties. We emphasize that this NFMG constitutes a unique material at the nanoscale level which is both magnetic and 100% organic. Both characteristics together give the NFMG an enormous plethora of potencial uses, ranging from applications in biomedicine (such as imaging, magnetic drug targeting and delivering, hyperthermia, etc) to applications in paints and inks.

The chemically modified magnetic graphite (MG) reported here was produced by a vapor phase redox reaction in closed nitrogen atmosphere ($N_2$, 1 atm.) with copper oxide using synthetic graphite powder (Fluka, granularity <0.1 mm). After obtaining the MG, we have prepared the nanofluid suspension (NFMG) by dissolving graphite in acetone, adding a Cetyltrimethylammonium bromide (CTAB) cationic surfactant, and bringing it to an ultra sonic edge. The resulting homogeneous solution was separated in a centrifuge at 6000 rpm. After five consequitive washes with acetone (to remove an excessive surfactant), deionized water was added and the solid sample was brought back to an ultra sonic edge for 1 min. The above procedure provided the necessary homogeneity and stability of the obtained solution. The adsorption of active agents on the surface of the graphite results from the favorable interaction between the surface and species of the solid adsorbents. Various interactions (such as electrostatic attraction, covalent binding, hydrogen binding, non-polar interactions and lateral interactions between adsorbed species) can contribute to the adsorption processes, facilitating the aqueous suspension of graphite. Recently, this field of research has been renewed by studies with fluid carbon nanotubes, among other forms.[15-17] To assess the presence of CTAB molecules on the surface of NFMG, we performed micro Raman analysis with samples of graphite dried in vacuum at a temperature of 60 °C for 6 hours.

The results (shown in Fig.1) revealed the presence of the peak at 1650 cm$^{-1}$ corresponding to NFMG, in addition to the peak at 1350 cm$^{-1}$ (known as the "disordered" D band) in the MG sample. For further comparison, Fig.1 also depicts the Raman spectrum showing characteristic bands of the surfactant used in the suspension (CTAB). Observe that, as expected, the CTAB functional groups, the hydrophobic (carbonic chain) and hydrophilic (cationic-active tensions with positive charge) parts of the surfactant correlate with the band structure of the graphite. Since the hydrophilic part tends to bind with the water molecules, its action results in stabilization of the nanofluid suspension. The structural characterization of NFMG was performed by transmission electron microscopy (TEM) using Philips CM-120 microscope. TEM analysis (see Fig.2) reveals a flake like morphology of NFMG. Relating the size of the scale in Fig.2 with the size of the particle in the nanofluid, the latter is estimated to be of the order of 10 nm. To test the magnetic properties of both MG and NFMG samples, we performed the standard zero field cooled (ZFC) measurements using a MPMS-5T SQUID magnetometer from Quantum Design. Fig.3 presents the effective ZFC curves for MG and NFMG samples (taken under the applied magnetic field of 1 kOe) after subtraction of the paramagnetic contributions. Notice that both MG and NFMG samples exhibit magnetization curves typical for ferromagnetic materials with the onset Curie temperatures around 300K. Thus, we can conclude that the aqueous suspension of graphite (NFMG) does not affect the room-temperature magnetization of the MG sample. Given an average value of 10nm for particle sizes in both MG and NFMG (deduced from TEM images), the low-temperature anomalies seen in Fig.3 for both samples are most likely related to the finite size effects. Besides, according to Fig.4 (which shows M-H curves at 2K and 300K), the hysteresis does not disappear with increasing the temperature and manifests itself in non-zero values of remnant magnetization and coercive magnetic field ($H_C$). Consequently, we can conclude that, even though the material has a relatively small value of $H_C$, it does exhibit a true ferromagnetic behavior up to 300K.

In summary, we have reported the synthesis, structural characterization and magnetic properties of nanofluid magnetic graphite (NFMG). The structural analysis of NFMG confirmed its stability in aqueous solution. By measuring the magnetization as a function of temperature and applied magnetic field in both MG and NFMG samples, we observed the typical ferromagnetic behavior. The comparative study unambiguously demonstrated that, after the chemical treatment, both MG and all its suspensions (prepared with acetone, CTAB and water) exhibit a stable net magnetization at room temperature.

This work has been financially supported by the Brazilian agencies CNPq, CAPES and FAPESP.

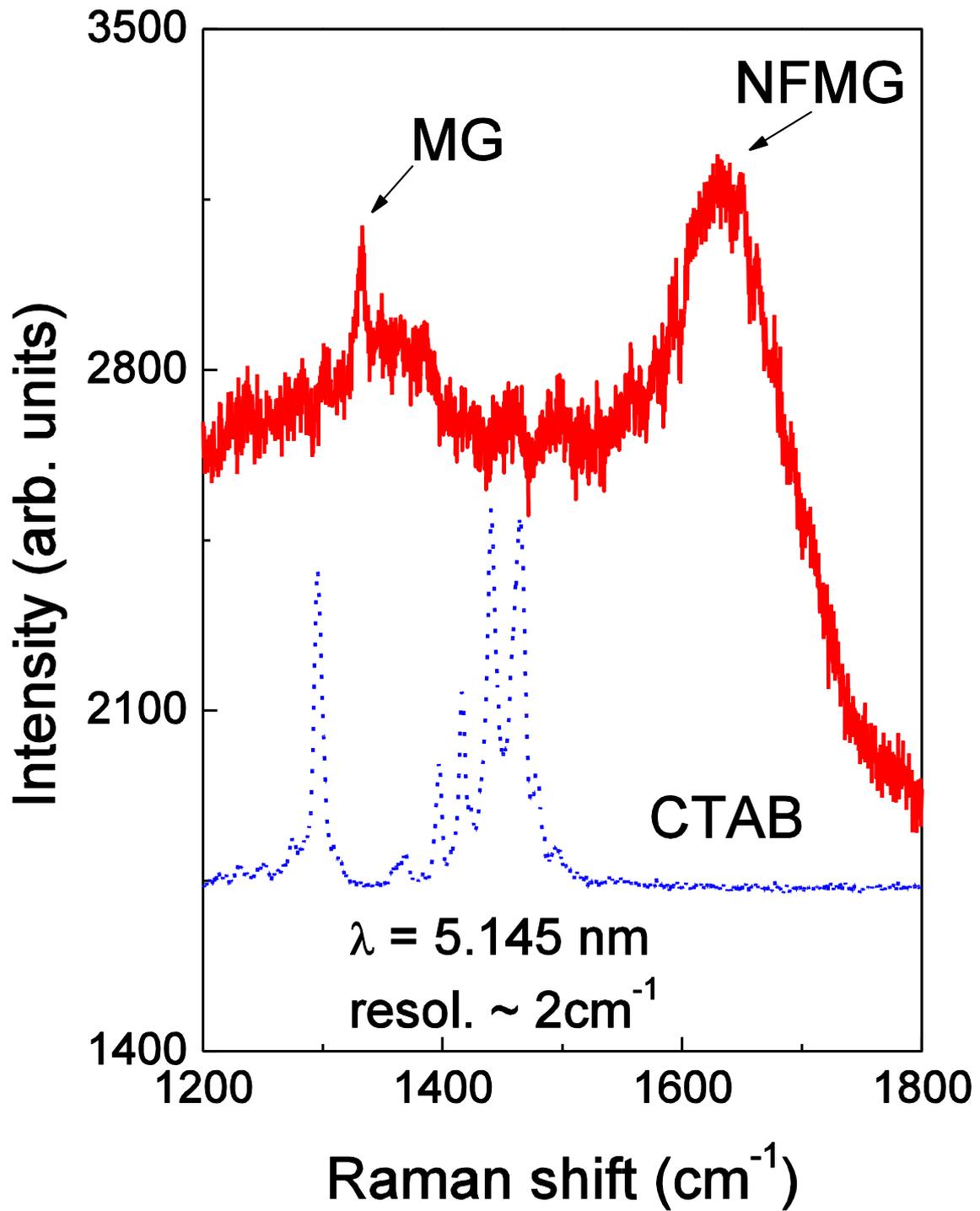

**Fig.1.** Raman spectra for chemically modified magnetic graphite (MG), surfactant used in the suspension (CTAB), and the aqueous suspension of the nanofluid magnetic graphite (NFMG).

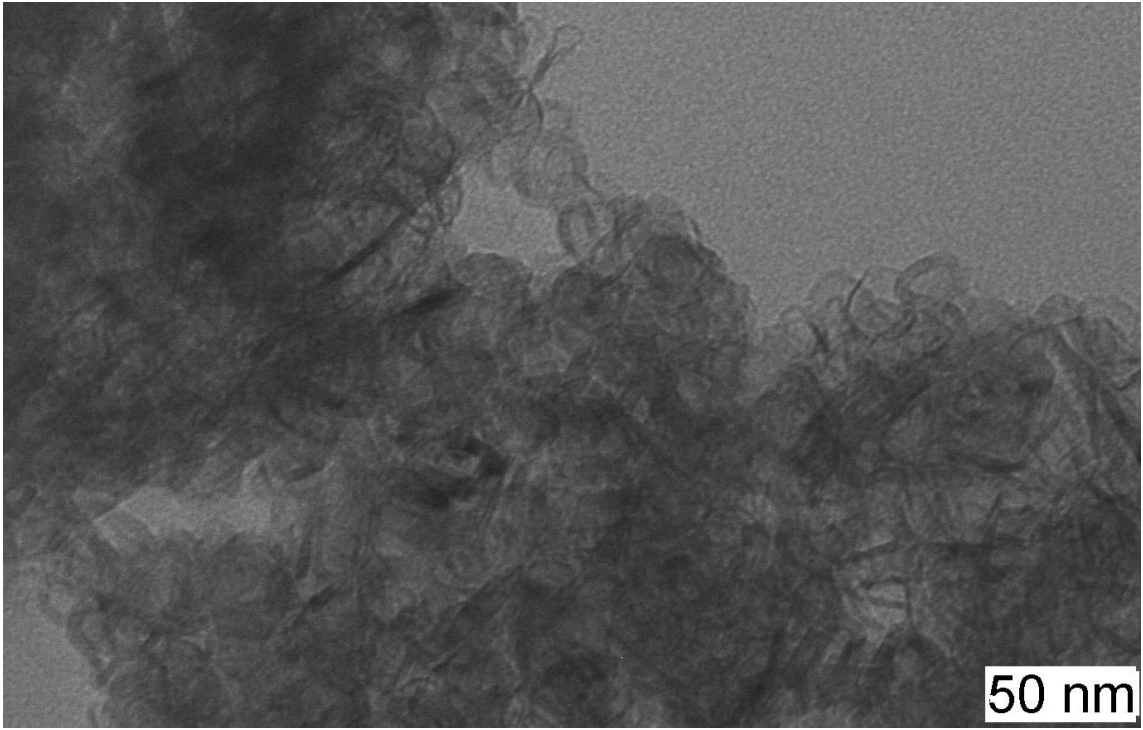

**Fig.2.** TEM image of NFMG sample showing a flake-like structure with an average size of the particle of the order of 10nm.

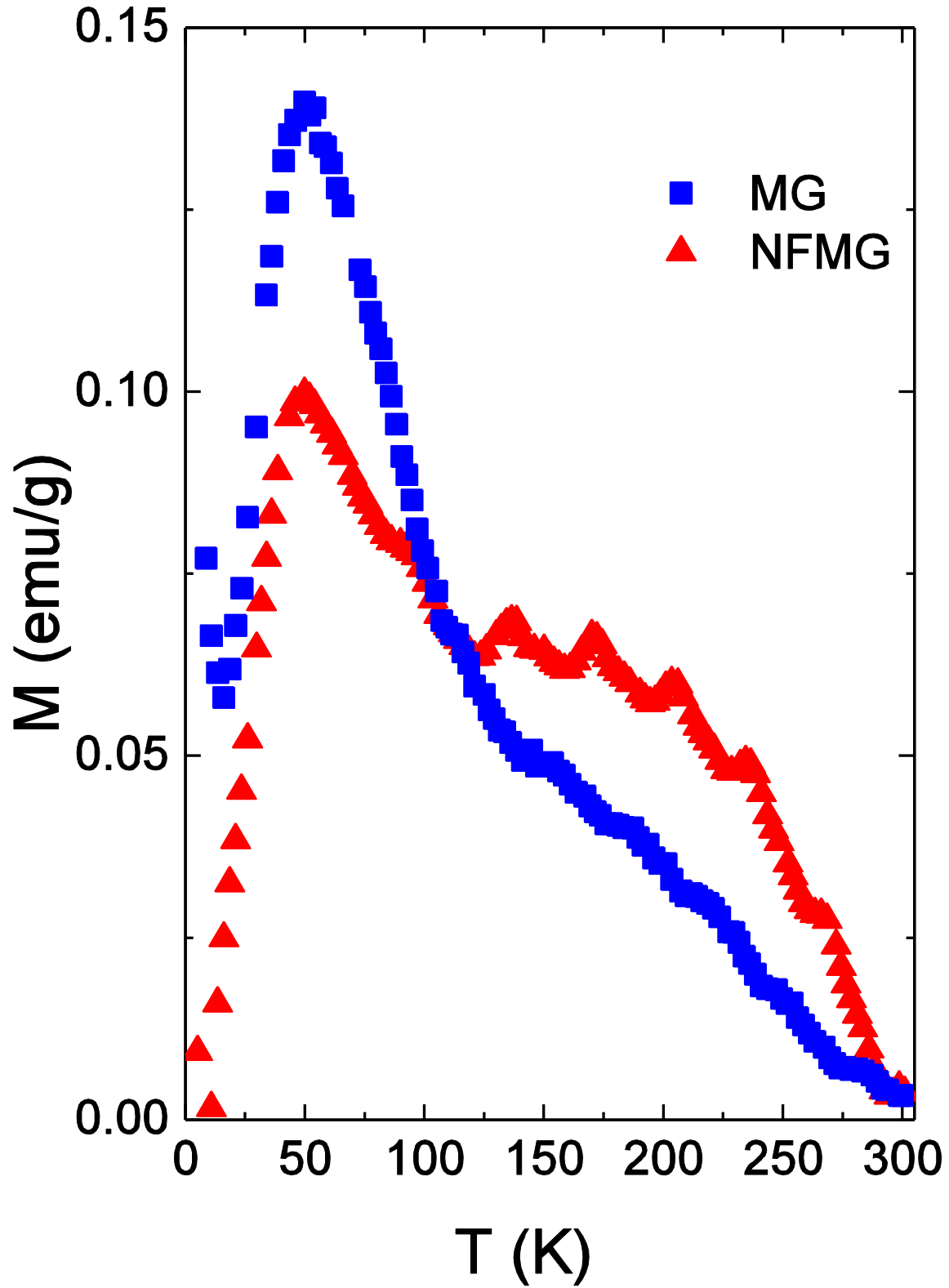

**Fig.3.** The temperature dependence of the effective ZFC magnetization for MG and NFMG samples (after subtracting the paramagnetic contributions).

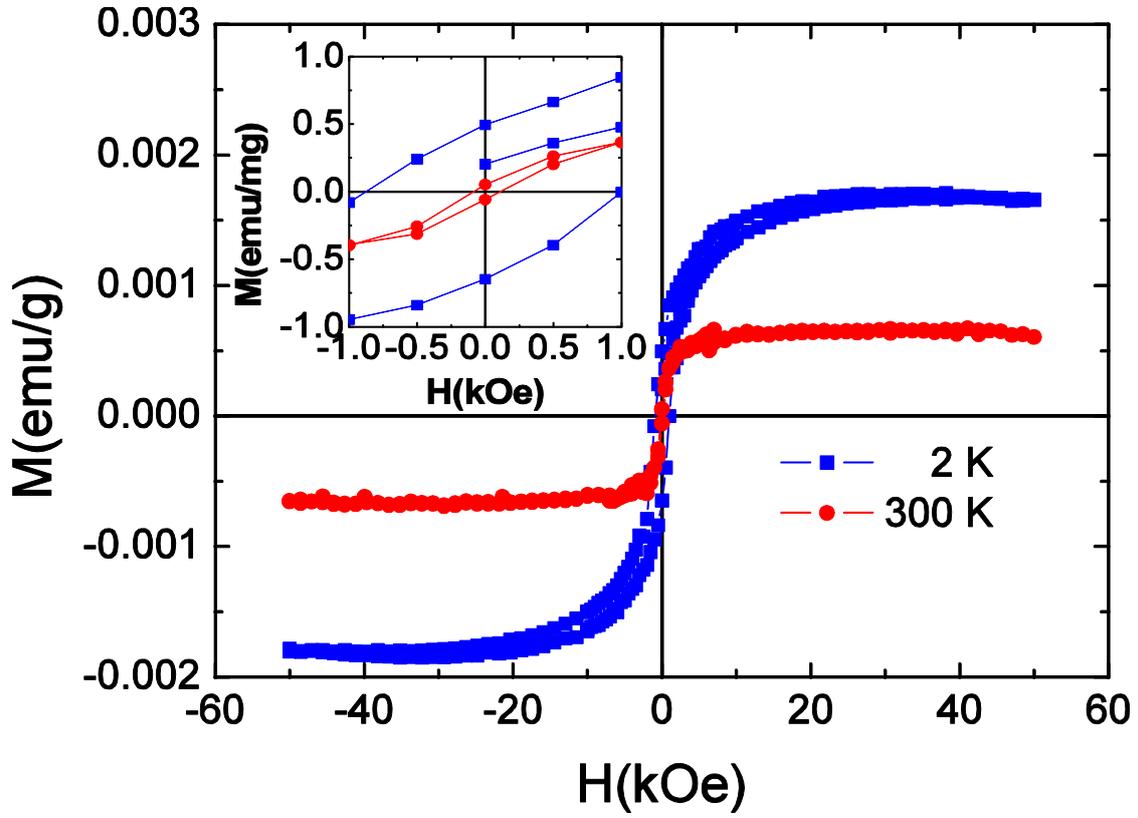

**Fig.4.** The hysteresis curves for NFMG sample for two temperatures (2K and 300K) showing a ferromagnetic like behavior of the sample. Inset: low-field M-H curves.